# Intermediate Excited State Relaxation Dynamics of Boron Vacancy Spin Defects in Hexagonal Boron Nitride

Paul Konrad[1], Mehran Kianinia[2], Lesley Spencer[2], Andreas Sperlich[1], Lukas Hein[1], Selin Steinicke[1], Igor Aharonovich[2], Vladimir Dyakonov[1*]

[1] Experimental Physics 6 and Würzburg-Dresden Cluster of Excellence ct.qmat, Julius-Maximilians-Universität Würzburg, Germany.
[2] School of Mathematics and Physical Sciences, University of Technology Sydney, Ultimo, NSW 2007, Australia.

* Corresponding author: vladimir.dyakonov@uni-wuerzburg.de

## Abstract

Optically addressable spin defects in hexagonal boron nitride offer promising potential for 2D quantum sensing, though excited-state dynamics remain poorly understood. In particular, the non-radiative relaxation paths from the excited triplet states to the ground state, especially those involving a shelving intermediate state (IS), remain largely hypothetical, and the rate constants have yet to be directly measured. In this work, we investigate the relaxation dynamics of the IS in the optical pumping cycle in a broad temperature range. We measure a 24.0(3) ns relaxation time from IS to the ground state at room temperature, which approximately doubles at low temperatures. Simulations reveal how spin populations and ground-state polarization evolve with varying excitation rate. Accordingly, we optimize optically detected magnetic resonance pulse sequences to account for the effects of IS relaxation. This considerably enhances spin manipulation efficiency, allowing substantial optimization of the quantum sensor's sensitivity based on boron vacancies.

## Introduction

Optically addressable solid-state spin defects have long been of interest since they offer a platform for quantum computing (*1*) and atomic scale sensing. (*2*) To date, room temperature spin-carrying defects such as nitrogen-vacancy centers in diamond, and defects in silicon carbide have been realized offering a material platform for quantum sensing. (*3–7*) The advantage of the intrinsic 3D nature of the host materials is that this protects the spin system from the environment. However, at the same time, this makes it difficult to use it as a sensor, as the imposed distance from sensor to the samples under investigation reduces the spin-spin interaction, while the proximity of surface defect states impairs the stability and spin coherence. For that matter, hexagonal boron nitride (hBN) is a promising candidate as the host material belongs to the class of 2D van-der-Waals materials. Compared to 3D host crystals, it has the decisive advantage that the position of the spin defect can be determined with atomic precision within a very thin layer, resulting in much shorter distances to the objects to be sensed.

Furthermore, hBN is an outstanding substrate material for developing various heterostructures with other 2D materials. (*8–10*) Its large bandgap of 6 eV provides electrical insulation but also offers a natural platform for intrinsic defects including room-temperature single photon emitters and various types of spin-carrying defects. (*11–16*) The negatively charged boron vacancy $V_B^-$ in hBN is a ground state spin triplet (*14, 17*) and holds a promising approach to circumvent the limitations of 3D materials. The spin states of the triplet and its surrounding nuclei are optically accessible and can be coherently controlled. (*18, 19*) Various applications have been successfully shown, including temperature, pressure, nuclear spin and magnetic field sensing. (*20–26*) The existence of these defects in few layer materials has also been demonstrated successfully. (*27*)
While the recent discovery of $V_B^-$ has sparked interest in utilizing these defects for sensing applications, research has mainly focused on photoluminescence (PL) quantum yields and spin-coherence properties, but essential parts of photo-dynamics are yet to be determined. Although there have been theoretical predictions of electronic level structure and associated transition rates (*28*), only few of the intrinsic rates have been determined experimentally. (*14, 29*) Only recently, pulsed PL measurements have been performed to fit a 5-level rate model and predict the transition rates. (*30–32*) By fitting the rate model, a lifetime of 30 ns was predicted for the metastable intermediate state (IS) of $V_B^-$ in hBN. The rate model fits have also been used for cryogenic temperature measurements. (*31*) While this approach provides insights into the transition rates between different electronic states of $V_B^-$, direct experimental measurement of the rates is essential to validate the model fits which rely on many unknown rates. This work deals with the direct experimental measurement of the intersystem crossing rate (ISC) from the IS back to the ground state (GS) and its influence on the spin relaxation dynamics as well as the limits of coherent control of $V_B^-$ spin centers in hBN.
In this study, we use a confocal microscope with a high-power (300 mW) continuous wave 473 nm diode laser that has fast on-off modulation capabilities (<2.5 ns rise time) together with time-correlated single-photon counting (TCSPC) to measure the PL dynamics as well as continuous wave and pulsed optically detected magnetic resonance (ODMR). For PL time traces upon switching the laser on, an initial overshoot in intensity is observed that depends on the duty cycle of laser excitation. We relate it to the recovery of GS population by the relaxation of shelved electrons from the IS during periods of no excitation. The IS relaxation is also observed in coherent control pulse sequences, when accounting for the relaxation by introducing a deliberate delay between laser turn-off and microwave manipulation. This notably increases manipulation effectiveness, which is shown by an increase in amplitude of Rabi oscillations and in flip operations of GS spin-polarization by microwave π-pulse.

## Results

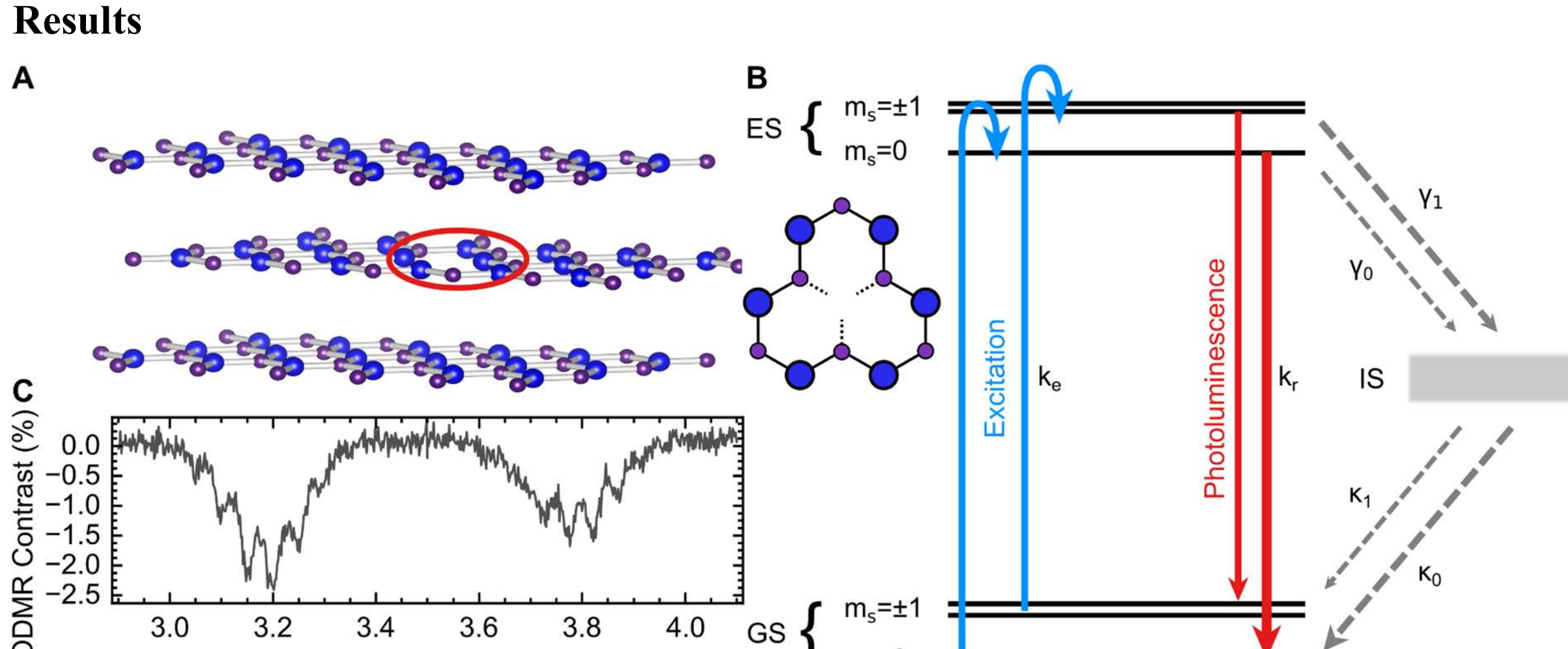

**Fig. 1**. **Signatures of the negatively charged boron vacancy.** (**A**) Left: Schematic of a boron vacancy defect in the middle layer of a multilayer hBN flake. Right: Top view of a hBN single layer with a boron vacancy. (**B**) Electronic structure of the negatively charged boron vacancy related $V_B^-$ spin triplet, which comprises ground (GS), excited (ES) and (metastable) intermediate states (IS). Assuming symmetric $m_s = \pm1$ spin sub-states, the system and its dynamics should be described using a 5-level rate model. (**C**) Continuous wave optically detected magnetic resonance (cwODMR) signal showing the microwave-induced transition from $m_s = 0$ to -1 at ~3.2 GHz and to $m_s = +1$ at ~3.8 GHz. Sub-structure arises from hyperfine interactions of spins with the three nearest neighbor nitrogen nuclei surrounding the boron vacancy.

## Energy Levels of $V_B^-$ Spin Triplet

Fig. 1a schematically shows a hBN layer with a $V_B^-$ defect embedded in between two defect-free hBN layers. Such an arrangement can be prototypical for sensing applications of $V_B^-$ in van der Waals heterostructures. The defects can be generated using various types of irradiation. (*14, 33–35*) In this work, we used a 30 keV nitrogen ion beam with a fluence of $\sim10^{14}$ ions/cm$^2$ to create defects in an hBN flake. The ground state of this defect is a spin triplet which can be addressed optically by laser excitation in the visible spectrum and detection of broad PL in the near infrared around 850 nm. The modeled electronic level structure of the defect is shown in Fig. 1b, containing a non-degenerate triplet GS with separated energy levels for the spin projection number $m_s = 0$, $m_s = -1$ and $m_s = +1$, a triplet excited state (ES) with the same degeneracy and an IS that is assumed to be a singlet. The GS zero-field splitting of the $m_s = 0$ and the $m_s = \pm1$ sub-levels is $D_{GS} = 3.49$ GHz and for the ES $D_{ES} = 2.09$ GHz (*29*). For simplification, the sub-manifolds of $m_s = \pm1$ of GS and ES are considered as one energy state, respectively. This reduces the rate model to an effective 5-level system.

The kinetics of the system are governed by the rates depicted in Fig. 1b. Excitation from the GS into the ES is spin independent and given by the spin-conserving rate $k_e$, whereas the radiative relaxation is given by the rate $k_r$. The spin dependent ISC transitions from ES to IS are given by the rates $\gamma_0, \gamma_1$ and from the IS to the GS by $\kappa_0, \kappa_1$ for the $m_s = 0$ and $m_s = \pm1$, respectively (see Fig. 1b).

Upon excitation of the GS, an emission around 850 nm can be observed and its intensity depends on the GS spin polarization. Spins excited from the $m_s = \pm1$ manifold are less likely to decay radiatively compared to the excited population of the $m_s = 0$ state. This is due to differences in the non-radiative relaxation rates $\gamma_0 < \gamma_1$ from the triplet ES into the IS. Further, an imbalance in the rates of transitions from the IS to the GS $\kappa_0 > \kappa_1$ leads to a preferential relaxation, i.e. GS spin-polarization into the $m_s$=0 state under continuous excitation. Such polarization can be modulated by resonant microwaves that match the GS ZFS splitting. By sweeping the microwave frequency at a constant magnetic field, a microwave-driven change in spin polarization at the resonant frequency is converted into a change in PL brightness (ODMR). Such a measurement is shown in Fig. 1c, where the transition frequencies from $m_s$ = 0 to $m_s$ = -1 and to $m_s$ = +1 can be seen as corresponding drops in photoluminescence. The signal is split 7-fold by additional resonances due to hyperfine interaction with the three surrounding $^{14}$N $I = 1$ nitrogen nuclei.

The IS plays a crucial role in gaining information about spin-polarization by PL measurements and initialization of the spin system by spin-polarization of the GS. Hence, the relaxation rates to and from the metastable IS are important parameters when designing pulse sequences of lasers and microwaves to achieve the highest spin polarization possible.

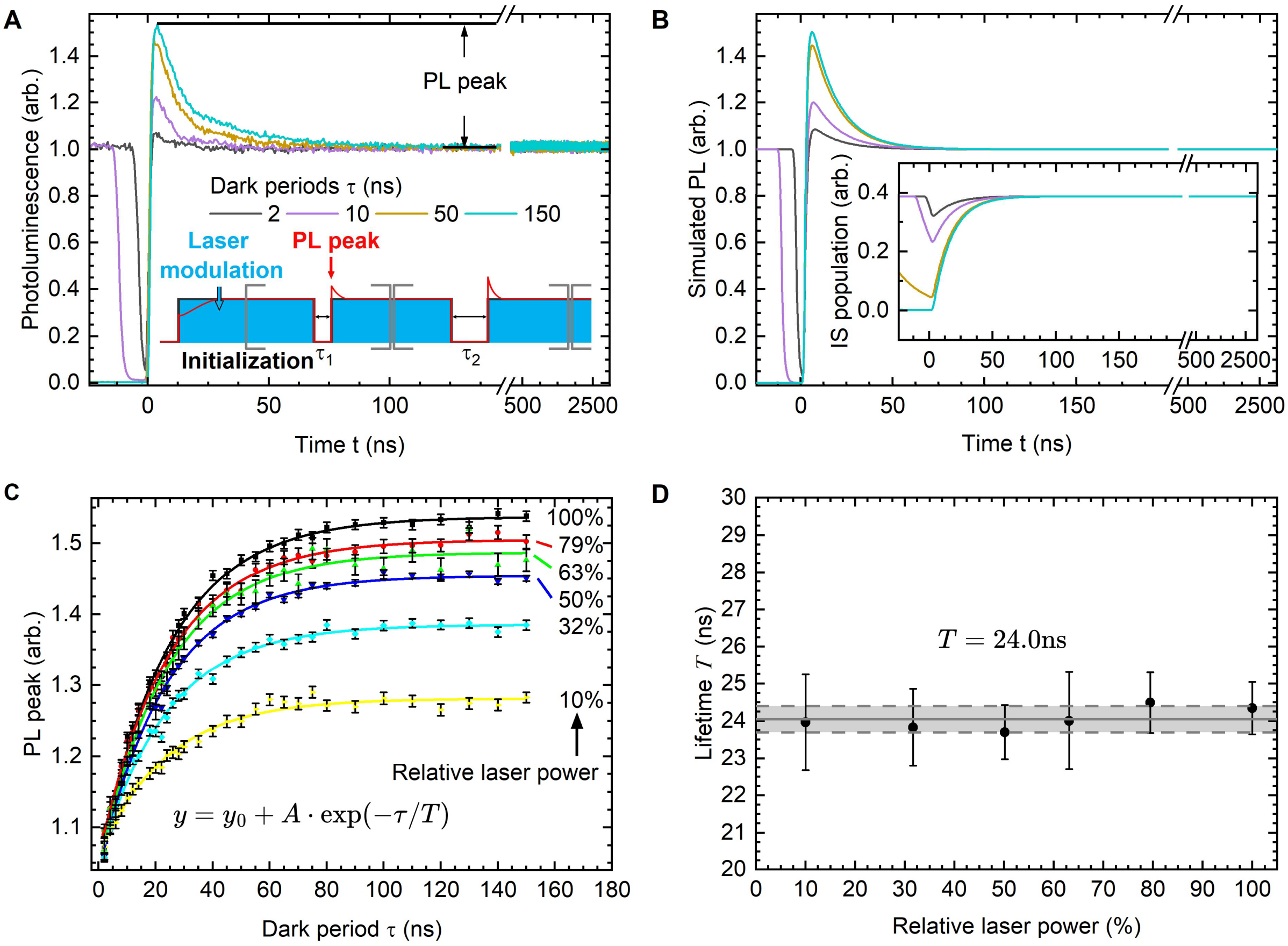

**Fig. 2. Time-resolved PL measurements. (A)** Transient PL for different dark periods τ. The initial PL overshoot is a clear indication of the release of shelved spins during the dark period. The transients then quickly reach a steady state. Inset: Pulse sequence necessary for probing the repopulation of the GS for a single dark time. The laser is turned off during the dark periods and switched on otherwise. Multiple measurements are chained together so the initialization pulse does not start from thermal equilibrium (**B**) Simulation of a 5-level model for the dark periods used in **(A)**. The experimental results are qualitatively confirmed, showing the same feature of an overshoot. Inset: Simulated population of the IS. Depending on the dark period, the state is partially depleted. (**C**) PL overshoot vs. the dark period for various laser powers showing exponential growth. The power at 100% is measured as 19mW before the objective lenses. (**D**) Lifetimes extracted from the fits in (c) with an average time constant of TIS = 24 ns in a wide range of applied laser powers.

## Ground State Repopulation and Intermediate State Lifetime

It is assumed that only the ES of the triplet system is optically active. This implies that decays (and lifetimes) of the other states involved in the optical pump cycle dynamics can only be detected *indirectly*. Their decay clearly influences the population of the ground state, as we shall see later. In our scenario, we consider only one IS, so that only this is responsible for the delayed dynamics of the excited spin states, thus acting as a shelving state.

To probe the relaxation of shelved spins, high excitation powers and sufficiently fast rising flanks of laser excitation are essential. These allow the excitation of a large portion of the GS population before fractions are shelved in the IS (see Supplementary Materials Fig. S1). In this way, a snapshot of the current population is shown in the beginning of the PL response to excitation. In Fig. 2a an all-optical measurement is conducted using two laser pulses, separated by a dark period $\tau$ (see inset). The first laser pulse ensures a predetermined, initialized state for all subsequent measurements. The second laser pulse is used to probe the current population by transient PL up to t = 3000 ns. The measurement is repeated with varying duty cycles of laser excitation as depicted in Fig. S2 and Fig. S3 with a composite pulse train. Therefore, laser read-out pulse from a measurement simultaneously serves as an initialization laser pulse for subsequent measurements (see Fig. S2). The PL response of the sample shows an initial overshoot that increases with a longer dark period $\tau$ (reduction in duty cycle). The ES was shown to decay rapidly in ~1.2 ns (*14*), so only the IS is still shelving electrons after the laser is turned off. The depletion of the IS population then leads to a repopulation of the GS during the dark period. This repopulation results in the observed PL overshoot, which subsequently decays during optical excitation due to reshelving of spins into the ES and IS. This method of probing dark state lifetimes is also referred to as fluorescence recovery and is known from e.g. NV diamond (*36*) , where the IS lifetime is in the order of 100 ns, and the ES has a substantially longer lifetime. Therefore, these measurements are experimentally not so demanding. A reference measurement of the substrate PL is shown in Fig. S4.

In our experiments, we used laser pulses of 3000 ns for initialization of the GS and varied the duty cycle by introducing a dark period from $\tau = 2$ ns to 150 ns between laser pulses as depicted in the Inset of Fig. 2a. Note that the dark period $\tau$ must not be confused with the time t marking the time tags of detected photons in time-resolved PL.

A selection of PL transients is shown in Fig. 2a, where the initial PL overshoot can be observed at the beginning of the read-out pulse. The overshoot decays rapidly and subsequently the PL remains constant, seen as a flat trace for t ≥ 500 ns. The PL overshoot intensity increases with the length of the dark period $\tau$. Note that all graphs are shown with the measurement laser pulse starting at time t = 0 to better compare the PL intensity for each dark period.

A simulation of the transient PL under pulsed excitation, considering the 5-level system with their respective decay rates, adequately reproduces the characteristic of the PL overshoot. The results are shown in Fig. 2b together with the simulated population of the IS in the inset. The full Python simulation code with used parameters is given in the Supplementary Materials.

To check the influence of the laser power, we repeat the measurement for different dark periods and various excitation powers. Here, 100% of laser power is equivalent to approximately 19mW at the objective lens aperture. The count rates were set to be around 1Mcps or less to avoid saturation effects of the single photon counter. To derive the relaxation rate from IS to GS, in Fig. 2c, the maximum PL overshoot is normalized to the steady state PL intensity for t ≥ 2000 ns and plotted against the corresponding dark period $\tau$. The figure also shows the same measurement for various excitation powers, spanning a range of an order of magnitude (10% - 100%). The data in Fig. 2c shows an exponential growth of the PL overshoot intensity which reaches a steady state value at $\tau$ > 150 ns. The data can be fitted with an exponential function:

$$f(\tau) = y_0 + A \cdot e^{-\tau/T_{IS}} \qquad (1)$$

yielding a time constant of $T_{\text{IS}}$ = 24.0(3) ns. The same value was obtained by fitting the results from different laser excitation powers as shown in Fig. 2d. This is in the same order of magnitude as the results of Whitefield et al. (*30*), who estimated IS lifetimes of ≈30 ns for $V_B^-$ in hBN by fitting time-resolved PL data to a 5-level model, as well as Clua-Provost et al. (*31*) with ≈18 ns.(*30, 31*) In our work, however, we use a direct approach to *measure* the IS lifetime which does not depend on a rate model. This is possible only if a high laser intensity in the saturation regime of the boron vacancy PL (see Fig. S5) with sufficiently steep rising flanks is used to probe the GS population in $V_B^-$. This is also shown in the simulation for slower response times of the laser, where the initial peak intensity vanishes and would be lost in the noise of a measurement (see Fig. S1).

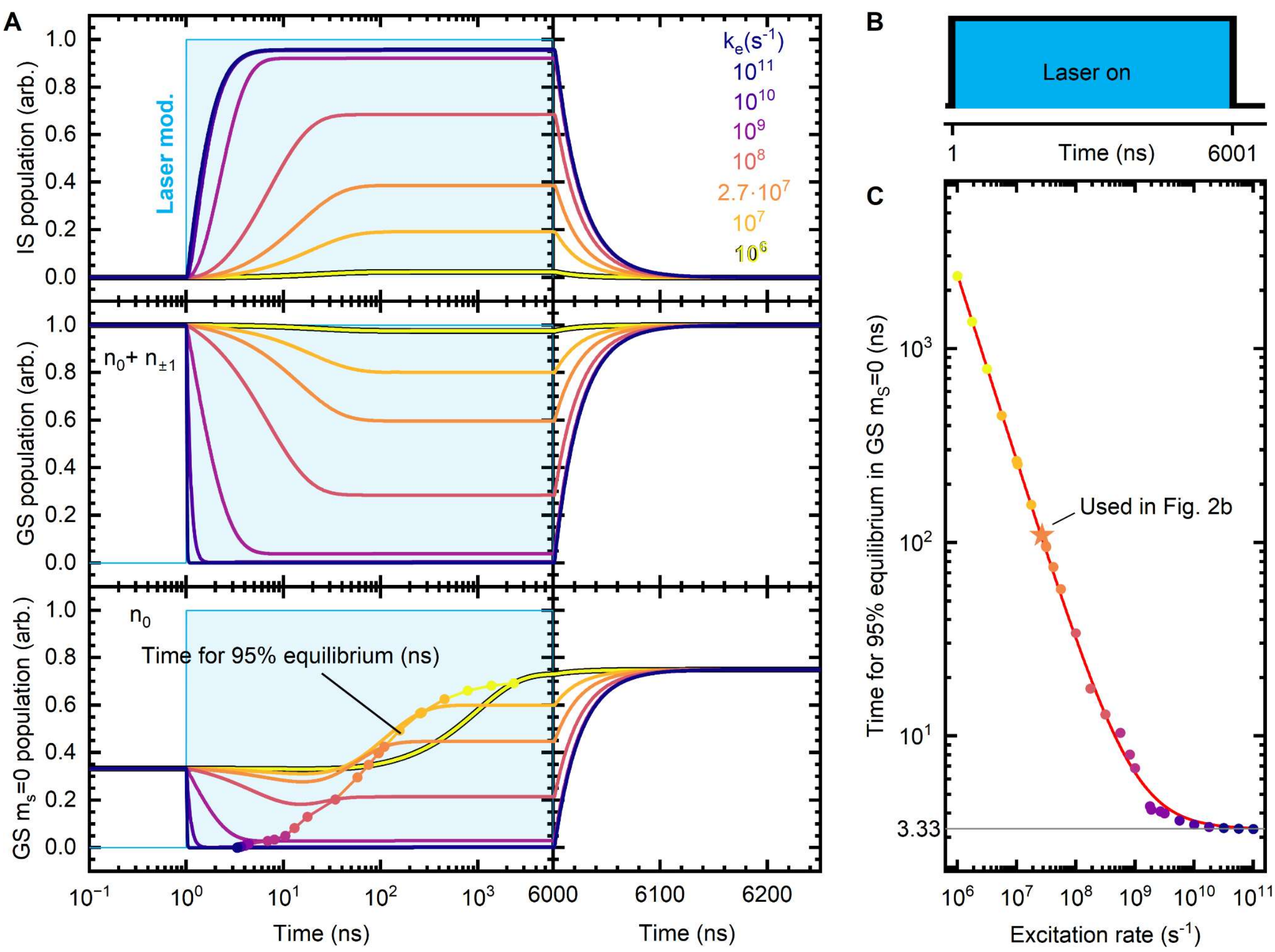

**Fig. 3. Simulation of the populations in the five-level system.** Populations are normalized to the total population. The laser is turned on for 6000 ns at time t = 1 ns. (**A**) Top: IS population depending on the excitation rate. Mid: Total population of the GS spin manifold ($n_0$+$n_{\pm1}$). Bottom: Population of the GS $m_s$ = 0 ($n_0$) spin-state. Eventually, a steady state is reached where the populations do not change anymore. At the highlighted points, 95% of the steady state population are reached (**B**) Simulated laser pulse sequence with turn-on at t = 1 ns and turn-off at t = 6001 ns. (**C**) Required laser turn-on time to achieve 95% equilibrium state population of the GS $m_s$ = 0 state. The data is taken from the points highlighted in (a) bottom. The red trace is a power law fit for excitation rates up to $10^{11}$ $s^{-1}$. The fastest time observed is $t_0$ = 3.33 ns and marked by a grey horizontal line.

Additionally, the simulation reveals the evolution of the five sub-level populations during optical excitation, starting from thermal equlibrium as shown in Fig. 3. Here, the laser is turned on for 6000 ns at time t = 1 ns, shown as the blue shaded area in Fig. 3a, for various excitation rates (laser power). The steady state reached at the end of the laser pulse at 6000ns marks the initialized state that is used as a starting condition for all measurements in this paper. The excitation rate used for the simulation of the experiment in Fig. 2b is $k_{exp}$ = 2.7·$10^7$ $s^{-1}$ is shown in orange. The top panel of Fig. 3a shows the evolution of the IS population. Remarkable is the increasing percentage of spins shelved in this state with increasing laser power up to ~96% of the total population. For $k_{exp}$ the simulation yields 39%. This is also reflected in the depopulation of the GS, shown in the mid panel of Fig. 2a. After laser turn-off, the repopulation of the GS occurs with a time constant of 24 ns (see previous discussion). Experimentally relevant is both the total population of the GS and the ratio between the $m_s$ = 0 state and $m_s$ = ±1 state $n_0$/$n_\pm$1, giving the spin polarization. The initial condition in the simulation is a population relation of 2:1 for the $m_s$ = ±1 state against $m_s$ = 0, corresponding to thermal equilibrium at room temperature. This leads to an initial population $n_0$ of 1/3 (normalized to the total number of spins), that is increased by excitation. After laser turn-on, an equilibrium state during laser excitation determined by the relation $\frac{\kappa_0}{\kappa_1}$ is eventually reached. The time needed to achieve a 95% of this equilibrium state is shown in Fig. 3c against the excitation rate. In the equilibrium state, the initialization of the system is complete and therefore gives the minimal pump time to prepare the system for ODMR. A power law ($t_{95\%} = t_0 + b \cdot k_e^a$) was fitted to the data yielding an exponent of $a$ = -0.95, shown in Fig. 3c as the red trace. This reflects the expected linear dependence of the initialization time on laser power for low powers. When the excitation rate is small compared to other system rates, the system dynamics are dominated by the excitation rate. For high rates, a saturation is observed with a minimal value of 3.33 ns (so 2.33 ns after laser turn-on) for 95% equilibrium. This value is shown as a grey horizontal line. All sub state populations with the threshold are shown in Fig. S6. The value of $t_{95\%}$ = 109 ns for $k_{exp}$ is highlighted as a star. Note that this value only indicates the polarization of the GS population and that shelved spins still need to be released by an additional ≈150 ns dark period (equal to $\sim 6 \times T_{IS}$), shown in the right panels of Fig. 3a. It is noteworthy that the ratio $\frac{\kappa_0}{\kappa_1}$ ultimately determines the maximum polarization of the GS but is not relevant for the shape of the transients and the interpretation of the PL overshoot in Fig. 2 (see Fig. S7).

**Influences of Intermediate State Depletion on Pulsed ODMR**

Next, we investigated the spin dynamics using a resonant microwave pulse to see the influence of GS repopulation by the IS relaxation on the efficiency of coherent control of the spin system. In pulsed ODMR measurements, resonant microwave pulses corresponding to the energy gap between the spin sub-states $m_S = 0$ and $m_S = \pm 1$ are used to stimulate an exchange of populations between these states. The resonance frequency is extracted from the ODMR measurement in Fig. 1c. Such manipulation is carried out during the dark period after a laser pulse that has previously polarized (initialized) $V_B^-$ into the $m_S = 0$ state. During this dark period, the GS will be repopulated by a decaying IS with the spin-polarized electrons until the IS is completely depleted. If the microwave manipulation is performed before a substantial fraction of the electrons shelved in the IS is relaxed to GS, the added population is not or only partially affected by the microwave pulse (see also Fig. 3a bottom). This results in reduced ODMR contrast. We have verified this behavior by measuring Rabi oscillations with a deliberate varying delay time (referred to as a buffer) between the laser's falling flank and the microwave pulses. This buffer simply allows partial depletion of the IS before microwave manipulation. Additionally, a reference measurement without a microwave is recorded and the contrast is calculated by $\Delta PL/PL_{reference}$, where PL is proportional to the (integrated) number of photons collected in the measurement window. Rabi measurements were conducted by increasing the pulse length $\tau$ of the microwave pulses and recording the time-correlated PL response (see Inset Fig. 4a). The data is then integrated in a measurement window. The Rabi measurement was repeated, and the buffer was increased. The buffer is therefore playing the role of the dark period in Fig. 2a. Two Rabi measurements with a buffer of 5 ns and 150 ns and an integration window of 60 ns are shown in Fig. 4a. The data obtained was then fitted with a damped sinusoidal of the form:

$$A\, e^{-\frac{\tau}{T_{2,\rho}}} \cdot sin\left(\frac{2\pi}{T} \cdot (\tau - \tau_0)\right), \qquad (2)$$

With amplitude A, damping $T_{2,\rho}$ = 56(1) ns, oscillation period $T$ and offset $\tau_0$.

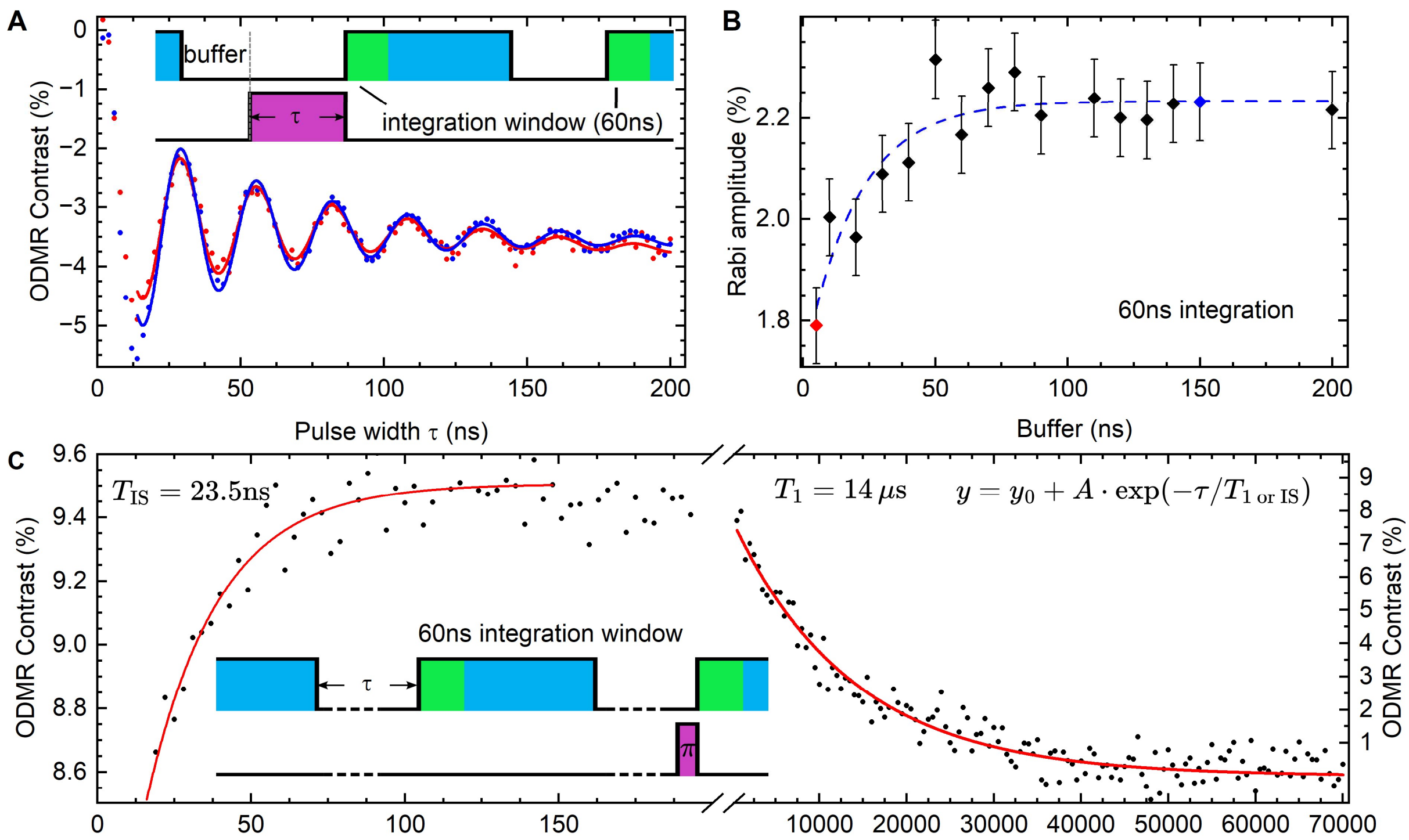

**Fig. 4. Coherent control of the $V_B^-$ spin system.** (**A**) Rabi oscillations of $V_B^-$ spin defects with a damped ($T_{2,\rho}$ = 56(1) ns) sinusoidal fit for buffer values of 5 ns (red) and 150 ns (blue). Inset: Pulse sequence with buffer, laser duty cycle (blue), microwave duty cycle (purple) and integration window (green). (**B**) Amplitude of the fit to the Rabi data over the buffer time before microwave manipulation. The data points related to the Rabi oscillations depicted in (a) are highlighted in the respective color. The buffer allows for the IS to release spins back into the GS, therefore increasing the Rabi amplitude. (**C**) Measurements of the spin-lattice relaxation time $T_1$, where a reference measurement with an applied $\pi$-pulse is added. Inset: Pulse sequence for measurements of $T_1$ with a dark period $\tau$ and reference measurement with an applied $\pi$-pulse to invert the spin polarization. Left: Measurement on the timescale of the IS lifetime. The data shows an increase in contrast with $T_{IS}$ = 23.5 ns, indicating an improved effect of the $\pi$-pulse. Right: Measurement on the timescale of the spin-lattice relaxation showing the decrease in polarization of the GS, approaching a steady state value in thermal equilibrium. An exponential decay is fitted to the data, yielding $T_1$ = 14 μs.

The amplitude A of the fit function is plotted in Fig. 4b against the buffer (dark period) applied. The Rabi amplitude increases as the buffer increases. In this experiment, the highest contrast was achieved for dark periods longer than ≈150 ns when the IS is completely depleted.

Finally, the effect of the IS relaxation rate can also be observed when measuring the spin-lattice relaxation time ($T_1$) of $V_B^-$ in the GS. In these measurements, the pulse sequence shown in the inset of Fig. 4c is applied, in which the length of the microwave pulse is fixed and set equal to the length of a $\pi$-pulse obtained from the Rabi measurement. Then, the buffer, i.e. the dark period time, is swept. It is like the measurement sequence shown in Fig. 2a with an additional reference measurement with a $\pi$-pulse. The microwave $\pi$-pulse flips the spin-polarization of initialized $V_B^-$ from the $m_s$ = 0 into $m_s$ = -1 sub-state, while the relaxation of IS continuously adds spin-polarized electrons to the GS. The measurements are performed for dark periods of different lengths. Exponential growth can be observed up to 200 ns, followed by exponential decay on the longer time scales over microseconds. The individual exponential fits are performed for both time scales. The timescale up to 200 ns yields an increase in amplitude on the timescale of the IS lifetime, corroborating the results obtained by the Rabi measurements. The fit for microsecond timescales yields the spin-lattice relaxation time of $T_1$ = 14 μs. It is much longer than the observed GS repopulation time of 24 ns, supporting the claims that the influence of spin-lattice relaxation on the observed dynamics of the IS depletion are negligible.

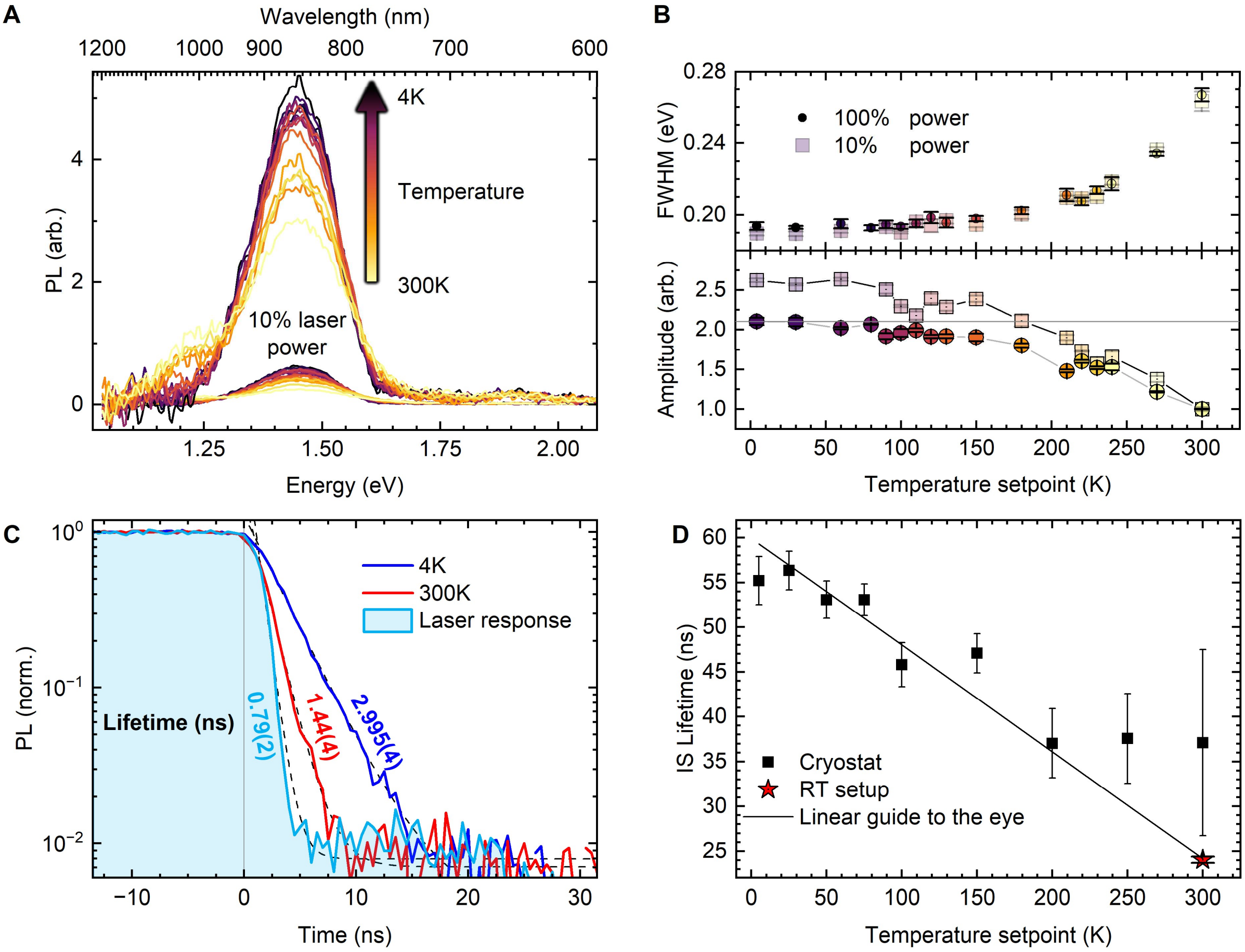

**Fig. 5. Temperature dependent PL measurements.** (**A**) PL spectra recorded at 100% laser power (upper traces) and 10% laser power (lower traces). The dominant $V_B^-$ PL peak around 850 nm is superimposing a weaker shoulder of PL from the silicon substrate at ≈1050 nm. (**B**) Full width at half maximum (top) of the $V_B^-$ PL peak and amplitude (bottom) for 100% and 10% laser power. The values were obtained by a double gaussian fit for $V_B^-$ and the silicon substrate PL. The amplitude values were normalized to the 300 K data point. (**C**) PL lifetime for temperatures of 4 K and 300 K together with the laser response. (**D**) Temperature dependence of the IS lifetime. Data points for T > 175 K are expected to be less accurate due to diminishing signal to noise ratio – with the exception of the room temperature data point measured separately in the RT setup. The data can be described by a linear fit as a guide to the eye.

These results also show that the effectiveness of the $\pi$-pulse in flipping the spin-polarization is increased, resulting in a greater ODMR contrast.

## Photoluminescence at cryogenic temperatures

Temperature dependent measurements were conducted in a closed cycle cryostat. The sample is placed on a copper pillar with a temperature sensor in its socket. The temperature is set with a resistive heating element. The spectra of $V_B^-$ show an increase in amplitude (see Fig. 5a and Fig. 5b) while the full width at half maximum FWHM decreases with decreasing temperature (see Fig. 5b). To determine the influence of laser heating inside the cryostat, the spectra were also recorded with only 10% laser power (smaller spectra in Fig. 5a). Additional to the $V_B^-$ PL, the spectra contain a shoulder due to the silicon substrate and therefore are fitted with a double gaussian function. From the fits the FWHM and the amplitude of the $V_B^-$ PL were extracted and are shown in Fig. 5b. A clear decrease in FWHM with temperature can be seen that is very similar for both laser powers. Even at cryogenic temperatures the spectra appear of Gaussian shape and do not show signs of a phonon sideband or zero phonon line. The amplitude is normalized to the value at 300 K for each laser power and shows a notable difference in curvature / saturation from 100% to 10% laser power. This difference suggests substantial influence of laser heating at cryogenic temperatures. Comparing the saturation amplitude for 100% laser power with the corresponding value at 10% shows a large temperature equivalent difference of approx. 180 K. Figure 5c shows the falling flank of PL and modulated laser response at 4 K and 300 K temperature setpoints. The laser response shows a very steep falling flank with an exponential fall time of 0.79(2) ns. The PL at 300 K shows $\tau_{\mathrm{ES,300K}}$ = 1.44(4) ns and $\tau_{\mathrm{ES,4K}}$ = 2.995(4) ns at 4K. The room temperature value is very close to previously reported measurements with pulsed laser excitation (*14*). Although the local temperature of the sample is unlikely to be 4 K, the lifetime is only doubled approximately. This is in agreement with pulsed measurements confirming the expected increase of ES lifetime (*31, 37, 38*).

To measure the PL recovery, modulation of the laser for the cryogenic setup is done with an acousto-optical modulator (AOM) instead of direct modulation of the laser source. We estimate the AOM rise time to 13.6 ns. This substantial increase in rise time renders the data acquisition to be restricted to the highest available laser power to resolve the PL recovery (see Fig. S1) due to considerable higher demands on the signal to noise ratio. The resulting IS lifetime is shown in Fig. 5d over the temperature setpoint. Above T = 175 K, the measurement of the overshoot becomes prohibitively difficult to determine since the signal-to-noise ratio cannot resolve the overshoot properly. The laser heating is expected to be less prominent than reported in Fig. 5b due to additional losses inside the optical beam path. The laser power is estimated to be close to the 10% of maximum power. The intermediate state lifetime was recorded for various temperature setpoints showing an increase up to 56 ns which is considerably longer than previously reported (*31*). Since laser heating still must be considered, these results suggest only a lower bound of IS relaxation time for local temperatures of 4 K. We like to emphasize that the influence of laser heating on the IS lifetime measured in the room temperature setup was tested by the measurements shown in Fig. 2c and did not show any influence.

**Discussion**

In this work, we use all-optical transient photoluminescence to access one of the crucial rate constants of the multi-level spin system related to the negatively charged boron vacancy $V_B^-$. Although this spin defect in the 2D van der Waals system hBN has distinct advantages over 3D host materials, it unfortunately suffers from a 100% nuclear magnetic environment, which shortens the spin coherence time compared to 3D host systems. It is therefore particularly important to determine the optimum parameters for coherent control which includes the timings in pulsed measurements. We have focused here on the metastable intermediate state that is responsible for the spin-polarized recombination to the ground state, which in turn is the basis for the ODMR technique commonly used to initialize, manipulate and read out the spin state. This state is experimentally difficult to access due to its optical inactivity, but we succeeded to directly determine its lifetime of 24 ns at room temperature by monitoring the repopulation of the $V_B^-$ ground state. Assuming a 5-level model of the electronic structure, this time constant can be independently assigned to the intermediate state. Simulations reveal the evolution of the populations in the sub-states and we provide a direct guide on achievable spin polarization in dependence of applied laser power and initialization pulse length. Furthermore, measurements at cryogenic temperatures reveal an approximate doubling of the IS lifetime, the PL intensity and the PL lifetime when lowering the temperature from ambient to 4 K. The width of the spectra decreases, but they do not reveal any structure or zero phonon line. We apply this knowledge to optimize pulsed ODMR sequences to account for the intermediate state lifetime. More specifically, our measurements show an increased ODMR contrast when a buffer of 150 ns corresponding to the depletion of the intermediate state is used between the initialization laser pulse and the microwave manipulation. The increase in ODMR contrast is especially shown in the effectiveness of the $\pi$-pulse as seen in the left panel of Fig. 4c. Here we observe an absolute contrast increase achieved by the pi pulse of 2%, which corresponds to a relative increase of nearly 26%. The increase of 26% reflects an increase in the number of addressed spins in the ensemble. The sensitivity for ensembles scales with $1/\sqrt{N}$ where $N$ is the number of addressed spins, hence experiences an increase of about 11% (*39, 40*). This is important information for all sensing applications using pulsed ODMR techniques like dynamical decoupling. In these multi-pulse sensing schemes, the first pulse applied to the system usually defines the portion of the spin system that is used for sensing. We estimated the number of addressed defects to be $\approx 2 \cdot 10^5$ (see Supplementary Materials). Since the increased availability of spins due to IS depletion is an intrinsic effect, this consideration also applies to single defects. As seen in the simulations in Fig. 3a this will see increased importance for higher excitation rates (e.g. by pulsed laser excitation) since more spins get shelved in the IS. Our studies show how pulsed ODMR measurements can be designed more efficiently and thus improve the applicability of $V_B^-$ spin defects in hBN for quantum applications.

## Materials and Methods

### Sample preparation

High pressure and high temperature hBN bulk crystals were sourced from the National Institute for Materials Science (NIMS). These contain boron and nitrogen isotopes in natural abundance. Flakes were mechanically exfoliated using scotch tape and directly transferred onto ozone cleaned Si substrates with 285nm $SiO_2$ at 80°C. Samples were then annealed for 4hrs at 500°C in air to remove residue from the exfoliation process. To create boron vacancies, a FEI DB235 Dual Beam FIB/SEM microscope was used. Fluences for each irradiated area were calculated according to the equation: $number\ of\ scans \times \frac{\left(\frac{beam\ current}{1e}\right)}{area} \times time\ for\ one\ scan\ of\ the\ area.$ A nitrogen ion beam was used for irradiations with a set beam energy of 30kV and measured beam current of 0.5pA. The areas being irradiated are squares of 4μm × 4μm. The fluence was varied by changing the number of scans of one area. From a single scan of 4μm square giving a fluence of $1.44 \times 10^{11}$ $cm^{-2}$ to 8000 scans of 4μm square giving a fluence of $1.15\times 10^{15}$ (See Supplementary Materials, Fig. S8.). The thickness of the hBN flake is estimated to 100 nm by atomic force microscopy.

### Room temperature confocal ODMR

PL data is collected by a home-built confocal microscope. The 473 nm excitation laser Cobolt 06-MLD is guided through a 100x objective (Zeiss EC epiplan-neofluar) with a numerical aperture of 0.9 onto the sample, creating a diffraction-limited spot of the size <1 μm. The optical pump rate of $k_{exp}=2.7\cdot10^7$ $s^{-1}$ that is used in Fig. 2b corresponds to a laser power of ~19 mW before the objective. Different pump rates are achieved by attenuating the laser with neutral density filters. The laser has a built-in modulation capability with a specified rise time <2.5 ns and modulation capability of 150 MHz. For laser modulation, no AOM has been used. The PL generated by the sample is split from the laser excitation by a dichroic mirror in combination with edge pass filters and collected using a single photon detector (Excelitas SPCM-AQRH-14 FC) in combination with a time-correlated single-photon counting module TimeTagger 20 from Swabian Instruments. Microwaves for coherent control are provided by a Stanford Research Systems SG380 signal generator that is switched on/off by a Mini-Circuits ZASWA-2-50DRA+ and amplified by a Vectawave VBA 2060-25. The MW is delivered to the sample by a 100 μm wire next to the measurement spot. Pulse sequences (for modulation of laser and MW) are programmed in Python for the Pulse Streamer 8/2 from Swabian instruments. Control of setup and data acquisition are done with Python.

### Cryogenic confocal setup

PL data is collected by a home-built confocal microscope above an attocube attoDRY800 closed cycle cryostat. The 473 nm excitation laser Cobolt 06-MLD is guided through an external 100x objective (Mitutoyo M Plan APO NIR HR) with a numerical aperture of 0.7 onto the sample. For modulated measurements, the laser is guided through an acousto-optical modulator before the objective. Detection and modulation are done with the same equipment as the room temperature setup. Spectra were recorded using a Nireos Gemini interferometer in combination with the Excelitas SPCM-AQRH-14 FC single photon detector.

## References

1. E. Knill, Physics: Quantum computing. *Nature* **463**, 441–443 (2010).
2. C. L. Degen, F. Reinhard, P. Cappellaro, Quantum sensing. *Rev. Mod. Phys.* **89** (2017).
3. J. M. Taylor, P. Cappellaro, L. Childress, L. Jiang, D. Budker, P. R. Hemmer, A. Yacoby, R. Walsworth, M. D. Lukin, High-sensitivity diamond magnetometer with nanoscale resolution. *Nature Phys* **4**, 810–816 (2008).
4. G. Balasubramanian, I. Y. Chan, R. Kolesov, M. Al-Hmoud, J. Tisler, C. Shin, C. Kim, A. Wojcik, P. R. Hemmer, A. Krueger, T. Hanke, A. Leitenstorfer, R. Bratschitsch, F. Jelezko, J. Wrachtrup, Nanoscale imaging magnetometry with diamond spins under ambient conditions. *Nature* **455**, 648–651 (2008).
5. V. M. Acosta, E. Bauch, M. P. Ledbetter, A. Waxman, L.-S. Bouchard, D. Budker, Temperature dependence of the nitrogen-vacancy magnetic resonance in diamond. *Phys. Rev. Lett.* **104**, 70801 (2010).
6. M. W. Doherty, V. M. Acosta, A. Jarmola, M. S. J. Barson, N. B. Manson, D. Budker, L. C. L. Hollenberg, Temperature shifts of the resonances of the NV− center in diamond. *Phys. Rev. B* **90** (2014).
7. H. Kraus, V. A. Soltamov, F. Fuchs, D. Simin, A. Sperlich, P. G. Baranov, G. V. Astakhov, V. Dyakonov, Magnetic field and temperature sensing with atomic-scale spin defects in silicon carbide. *Scientific reports* **4**, 5303 (2014).
8. C. R. Dean, A. F. Young, I. Meric, C. Lee, L. Wang, S. Sorgenfrei, K. Watanabe, T. Taniguchi, P. Kim, K. L. Shepard, J. Hone, Boron nitride substrates for high-quality graphene electronics. *Nature nanotechnology* **5**, 722–726 (2010).
9. T. Georgiou, R. Jalil, B. D. Belle, L. Britnell, R. V. Gorbachev, S. V. Morozov, Y.-J. Kim, A. Gholinia, S. J. Haigh, O. Makarovsky, L. Eaves, L. A. Ponomarenko, A. K. Geim, K. S. Novoselov, A. Mishchenko, Vertical field-effect transistor based on graphene-WS2 heterostructures for flexible and transparent electronics. *Nature nanotechnology* **8**, 100–103 (2013).
10. A. V. Kretinin, Y. Cao, J. S. Tu, G. L. Yu, R. Jalil, K. S. Novoselov, S. J. Haigh, A. Gholinia, A. Mishchenko, M. Lozada, T. Georgiou, C. R. Woods, F. Withers, P. Blake, G. Eda, A. Wirsig, C. Hucho, K. Watanabe, T. Taniguchi, A. K. Geim, R. V. Gorbachev, Electronic properties of graphene encapsulated with different two-dimensional atomic crystals. *Nano letters* **14**, 3270–3276 (2014).
11. T. T. Tran, K. Bray, M. J. Ford, M. Toth, I. Aharonovich, Quantum emission from hexagonal boron nitride monolayers. *Nature nanotechnology* **11**, 37–41 (2016).
12. G. Grosso, H. Moon, B. Lienhard, S. Ali, D. K. Efetov, M. M. Furchi, P. Jarillo-Herrero, M. J. Ford, I. Aharonovich, D. Englund, Tunable and high-purity room temperature single-photon emission from atomic defects in hexagonal boron nitride. *Nature communications* **8**, 705 (2017).
13. M. Abdi, J.-P. Chou, A. Gali, M. B. Plenio, Color Centers in Hexagonal Boron Nitride Monolayers: A Group Theory and Ab Initio Analysis. *ACS Photonics* **5**, 1967–1976 (2018).
14. A. Gottscholl, M. Kianinia, V. Soltamov, S. Orlinskii, G. Mamin, C. Bradac, C. Kasper, K. Krambrock, A. Sperlich, M. Toth, I. Aharonovich, V. Dyakonov, Initialization and read-out of intrinsic spin defects in a van der Waals crystal at room temperature. *Nature materials* **19**, 540–545 (2020).
15. N. Chejanovsky, A. Mukherjee, J. Geng, Y.-C. Chen, Y. Kim, A. Denisenko, A. Finkler, T. Taniguchi, K. Watanabe, D. B. R. Dasari, P. Auburger, A. Gali, J. H. Smet, J. Wrachtrup, Single-spin resonance in a van der Waals embedded paramagnetic defect. *Nature materials* **20**, 1079–1084 (2021).

16. H. L. Stern, Q. Gu, J. Jarman, S. Eizagirre Barker, N. Mendelson, D. Chugh, S. Schott, H. H. Tan, H. Sirringhaus, I. Aharonovich, M. Atatüre, Room-temperature optically detected magnetic resonance of single defects in hexagonal boron nitride. *Nature communications* **13**, 618 (2022).
17. V. Ivády, G. Barcza, G. Thiering, S. Li, H. Hamdi, J.-P. Chou, Ö. Legeza, A. Gali, Ab initio theory of the negatively charged boron vacancy qubit in hexagonal boron nitride. *npj Comput Mater* **6** (2020).
18. A. Gottscholl, M. Diez, V. Soltamov, C. Kasper, A. Sperlich, M. Kianinia, C. Bradac, I. Aharonovich, V. Dyakonov, Room temperature coherent control of spin defects in hexagonal boron nitride. *Science advances* **7** (2021).
19. X. Gao, S. Vaidya, K. Li, P. Ju, B. Jiang, Z. Xu, A. E. L. Allcca, K. Shen, T. Taniguchi, K. Watanabe, S. A. Bhave, Y. P. Chen, Y. Ping, T. Li, Nuclear spin polarization and control in hexagonal boron nitride. *Nature materials* **21**, 1024–1028 (2022).
20. A. Gottscholl, M. Diez, V. Soltamov, C. Kasper, D. Krauße, A. Sperlich, M. Kianinia, C. Bradac, I. Aharonovich, V. Dyakonov, Spin defects in hBN as promising temperature, pressure and magnetic field quantum sensors. *Nature communications* **12**, 4480 (2021).
21. W. Liu, Z.-P. Li, Y.-Z. Yang, S. Yu, Y. Meng, Z.-A. Wang, Z.-C. Li, N.-J. Guo, F.-F. Yan, Q. Li, J.-F. Wang, J.-S. Xu, Y.-T. Wang, J.-S. Tang, C.-F. Li, G.-C. Guo, Temperature-Dependent Energy-Level Shifts of Spin Defects in Hexagonal Boron Nitride. *ACS Photonics* **8**, 1889–1895 (2021).
22. J.-P. Tetienne, Quantum sensors go flat. *Nature Phys* **17**, 1074–1075 (2021).
23. M. Huang, J. Zhou, Di Chen, H. Lu, N. J. McLaughlin, S. Li, M. Alghamdi, D. Djugba, J. Shi, H. Wang, C. R. Du, Wide field imaging of van der Waals ferromagnet Fe3GeTe2 by spin defects in hexagonal boron nitride. *Nature communications* **13**, 5369 (2022).
24. A. J. Healey, S. C. Scholten, T. Yang, J. A. Scott, G. J. Abrahams, I. O. Robertson, X. F. Hou, Y. F. Guo, S. Rahman, Y. Lu, M. Kianinia, I. Aharonovich, J.-P. Tetienne, Quantum microscopy with van der Waals heterostructures. *Nature Phys* **19**, 87–91 (2023).
25. R. Rizzato, M. Schalk, S. Mohr, J. C. Hermann, J. P. Leibold, F. Bruckmaier, G. Salvitti, C. Qian, P. Ji, G. V. Astakhov, U. Kentsch, M. Helm, A. V. Stier, J. J. Finley, D. B. Bucher, Extending the coherence of spin defects in hBN enables advanced qubit control and quantum sensing. *Nature communications* **14**, 5089 (2023).
26. C. J. Patrickson, S. Baber, B. B. Gaál, A. J. Ramsay, I. J. Luxmoore, High frequency magnetometry with an ensemble of spin qubits in hexagonal boron nitride. *npj Quantum Inf* **10** (2024).
27. A. Durand, T. Clua-Provost, F. Fabre, P. Kumar, J. Li, J. H. Edgar, P. Udvarhelyi, A. Gali, X. Marie, C. Robert, J. M. Gérard, B. Gil, G. Cassabois, V. Jacques, Optically Active Spin Defects in Few-Layer Thick Hexagonal Boron Nitride. *Phys. Rev. Lett.* **131**, 116902 (2023).
28. J. R. Reimers, J. Shen, M. Kianinia, C. Bradac, I. Aharonovich, M. J. Ford, P. Piecuch, Photoluminescence, photophysics, and photochemistry of the VB− defect in hexagonal boron nitride. *Phys. Rev. B* **102** (2020).
29. S. Baber, R. N. E. Malein, P. Khatri, P. S. Keatley, S. Guo, F. Withers, A. J. Ramsay, I. J. Luxmoore, Excited State Spectroscopy of Boron Vacancy Defects in Hexagonal Boron Nitride Using Time-Resolved Optically Detected Magnetic Resonance. *Nano letters* **22**, 461–467 (2022).
30. B. Whitefield, M. Toth, I. Aharonovich, J.-P. Tetienne, M. Kianinia, Magnetic Field Sensitivity Optimization of Negatively Charged Boron Vacancy Defects in hBN. *Adv Quantum Tech* (2023).
31. T. Clua-Provost, Z. Mu, A. Durand, C. Schrader, J. Happacher, J. Bocquel, P. Maletinsky, J. Fraunié, X. Marie, C. Robert, G. Seine, E. Janzen, J. H. Edgar, B. Gil, G. Cassabois, V.

Jacques, Spin-dependent photodynamics of boron-vacancy centers in hexagonal boron nitride. *Phys. Rev. B* **110** (2024).
32. W. Lee, V. S. Liu, Z. Zhang, S. Kim, R. Gong, Du Xinyi, K. Pham, T. Poirier, Z. Hao, J. H. Edgar, P. Kim, zu Chong, E. J. Davis, N. Y. Yao, "Intrinsic high-fidelity spin polarization of charged vacancies in hexagonal boron nitride". (17.06.2024; http://arxiv.org/pdf/2406.11953v1).
33. X. Gao, S. Pandey, M. Kianinia, J. Ahn, P. Ju, I. Aharonovich, N. Shivaram, T. Li, Femtosecond Laser Writing of Spin Defects in Hexagonal Boron Nitride. *ACS Photonics* **8**, 994–1000 (2021).
34. M. Kianinia, S. White, J. E. Fröch, C. Bradac, I. Aharonovich, Generation of Spin Defects in Hexagonal Boron Nitride. *ACS Photonics* **7**, 2147–2152 (2020).
35. F. F. Murzakhanov, B. V. Yavkin, G. V. Mamin, S. B. Orlinskii, I. E. Mumdzhi, I. N. Gracheva, B. F. Gabbasov, A. N. Smirnov, V. Y. Davydov, V. A. Soltamov, Creation of Negatively Charged Boron Vacancies in Hexagonal Boron Nitride Crystal by Electron Irradiation and Mechanism of Inhomogeneous Broadening of Boron Vacancy-Related Spin Resonance Lines. *Nanomaterials (Basel, Switzerland)* **11** (2021).
36. N. B. Manson, J. P. Harrison, M. J. Sellars, Nitrogen-vacancy center in diamond: Model of the electronic structure and associated dynamics. *Phys. Rev. B* **74** (2006).
37. Z. Mu, H. Cai, D. Chen, Z. Jiang, S. Ru, X. Lyu, X. Liu, I. Aharonovich, W. Gao, Excited-state optically detected magnetic resonance of spin defects in hexagonal boron nitride. *Phys. Rev. Lett.* **128** (2022).
38. N. Mendelson, R. Ritika, M. Kianinia, J. Scott, S. Kim, J. E. Fröch, C. Gazzana, M. Westerhausen, L. Xiao, S. S. Mohajerani, S. Strauf, M. Toth, I. Aharonovich, Z.-Q. Xu, Coupling Spin Defects in a Layered Material to Nanoscale Plasmonic Cavities. *Advanced materials (Deerfield Beach, Fla.)* **34**, e2106046 (2022).
39. J. F. Barry, J. M. Schloss, E. Bauch, M. J. Turner, C. A. Hart, L. M. Pham, R. L. Walsworth, Sensitivity optimization for NV-diamond magnetometry. *Rev. Mod. Phys.* **92** (2020).
40. D. Budker, M. Romalis, Optical magnetometry. *Nature Phys* **3**, 227–234 (2007).
41. V. M. Acosta, E. Bauch, M. P. Ledbetter, C. Santori, K.-M. C. Fu, P. E. Barclay, R. G. Beausoleil, H. Linget, J. F. Roch, F. Treussart, S. Chemerisov, W. Gawlik, D. Budker, Diamonds with a high density of nitrogen-vacancy centers for magnetometry applications. *Phys. Rev. B* **80** (2009).
42. A. Patra, P. Konrad, A. Sperlich, T. Biktagirov, W. G. Schmidt, L. Spencer, I. Aharonovich, S. Höfling, V. Dyakonov, Quantifying Spin Defect Density in hBN via Raman and Photoluminescence Analysis. *Adv. Funct. Mater.* (2025).

**Funding**

European Research Council (ERC) (Grant agreement No. 101055454) (VD)

Lighthouse project IQ-Sense of the Bavarian State Ministry of Science and the Arts as part of the Bavarian Quantum Initiative Munich Quantum Valley (15 02 TG 86). (PK)

Würzburg-Dresden Cluster of Excellence on Complexity and Topology in Quantum Matter ct.qmat (EXC 2147, DFG project ID 390858490) (AS, VD)

Australian Research Council (CE200100010) (IA, MK)

Asian Office of Aerospace Research and Development (FA2386-20-1-4014) for financial support. (IA, MK)

**Author Contributions**

Conceptualization: PK, AS, VD, IA, MK
Data curation: PK, SS, LH
Formal analysis: PK, SS, LH
Funding acquisition: VD
Investigation: PK, LH, SS, LS, MK
Methodology: PK, AS, LS, VD
Project Administration: VD, AS, IA
Resources: LS, IA, VD, AS
Supervision: AS, IA, VD, MK
Validation: VD, AS, IA, PK, SS, LH
Visualization: PK
Writing – original draft: PK, MK
Writing – review and editing: PK, MK, AS, VD, LS, IA

**Competing Interests**

All other authors declare they have no competing interests.

**Data and Materials Availability**

All data needed to evaluate the conclusions in the paper are present in the main paper and the Supplementary Materials. The Python code that supports the findings within this study is provided with this paper.

**Supplementary Materials**

- Supplementary Text
- Figs. S1 to S8
- Code S1
- Data S1

# Supplementary Materials for

## Intermediate Excited State Relaxation Dynamics of Boron-Vacancy Spin Defects in Hexagonal Boron Nitride

Paul Konrad *et al.*

*Corresponding author. Email: vladimir.dyakonov@uni-wuerzburg.de

**This PDF file includes:**

Supplementary Text
Figs. S1 to S8

**Other Supplementary Materials for this manuscript include the following:**

Code S1
Data S1

**Supplementary Text**

Sensitivity Estimation for Spin Packet

We estimate the spin-projection limited sensitivity $\delta B \approx \frac{h}{g_e \mu_B R \sqrt{\eta}\sqrt{N\tau T_2^*}}$ given by (20,41) for our current setup. Here, $h$ is the Planck constant, $g_e$ is the gyromagnetic ration, $\mu_B$ is the Bohr magneton, R is the ODMR contrast, $\eta$ is the collection efficiency, $N$ is the number of addressed defects, $\tau$ is the interrogation time and $T_2^*$ is the dephasing time. Unfortunately, we have no direct measure of the absolute number of spins we address since classical EPR methods fail due to the small sample volume (and inhomogeneity). A recent own publication allows us however to estimate the number of simultaneously addressed spins via Raman features (42) and we can assume to address $5.8 \cdot 10^{18} \frac{\text{defects}}{\text{cm}^3} * \pi * 1.22 * \left(\frac{473\text{nm}}{2\,NA}\right)^2 * 100\text{nm}$ , with a diffraction limited spot of size $\approx 1.22 * \frac{473\text{nm}}{NA}$, a numerical aperture of the objective of $NA = 0.9$ and a flake thickness of 100nm. These assumptions yield an estimate of $2 \cdot 10^5$ addressed defects. For measurement time we can also take $\tau = 1ns$. The collection efficiency $\eta$ can be estimated to be significantly higher than for the previous publication by Gottscholl et al., since the numerical aperture changed from 0.3 to 0.9. Nevertheless, for comparison and since a lot more factors determine the collection efficiency, we keep the assumption of the efficiency to be 1%. The Contrast of the ODMR measurement also significantly increased to 2.5% as shown in Fig. 1 of the main text. The decoherence time $T_2^*$ can be estimated by the envelope of the Rabi oscillation to 56ns (see main text). With these assumptions, one obtains a minimal detectable magnetic field of $\delta B \approx 100\text{nT}/\sqrt{\text{Hz}}$ which is in the same order of magnitude as from Gottscholl et al (20).

5-Level Rate Model:

In this work we use a simplified 5-level model for the $V_B^-$ spin system shown in Fig. 1(b) of the main text. The system is assumed to be split into a triplet sub-system and a singlet sub-system with a single metastable intermediate state (IS). The triplet sub-system has a non-degenerate ground and excited state (GS, ES) where the energy levels are associated with the projections of the spin $m_S = 0$ and $m_s = \pm 1$ (combined into one level) are split by the zero-field splitting of $D_{GS}$ = 3.49 GHz (*14*) and $D_{ES}$ = 2.09 GHz (*29*), respectively. Electrons can be excited by visible laser light with a relaxation path emitting photons in the near infrared around 850 nm (*14, 34*). From the ES, electrons can undergo spin-conserving optical relaxation or spin-selective and non-conserving relaxation via an intermediate state (*28*). The spin-selective relaxation enables optically detected magnetic resonance (ODMR) and the non-conserving property together with favored relaxation into the $m_s = 0$ ground state leads to spin polarization which also allows for coherent control (*18*).

Simulation code

The Python code provided as a separate file is used to simulate the PL response inside of the 5-level rate model. The excitation rate can be modified to account for the smearing of the laser power. Rising and settling of the excitation rate is simulated smoothly by a hyperbolic function. The PL of the 5-level model is simulated by the code submitted with this manuscript.

For simulations, the starting values for rates are adapted from literature:

$k_e = 27\ \mu s^{-1}$

$k_r = 90.9\ ms^{-1}$ (*28*)

$\gamma_0 = 820\ \mu s^{-1}$ (*31*)

$\gamma_1 = 1.89\ ns^{-1}$ (*31*)

$\kappa_0 = 31.2\ \mu s^{-1}$

$\kappa_1 = 1/3 \cdot \kappa_0$ (*29, 31*)

The value of $\kappa_0$ is determined by the relation $\kappa_1 = 1/3 \cdot \kappa_0$ and the IS lifetime $T_{IS} = 1/(\kappa_0+\kappa_1) = 24.0(3)$ ns. The values of $\gamma_0$ and $\gamma_1$ were determined in (*31*) using pulsed laser excitation. This is a method that we expect to be highly reliable. It is noteworthy that (*29*) used the same approach with shorter laser pulses, hence expected better instrument response function. However, as Clua-Provost et al. showed with 18 samples, there is slight deviation from sample to sample. We therefore use the statistically more robust value from (*31*).

The relation $\kappa_1 = 1/3 \cdot \kappa_0$ is well agreed upon between Clua-Provost et al. and Baber et al., although a different notation is used (*29, 31*). In the first, $\kappa_1$ is used for the relaxation rate from the intermediate state into a single ground state level (assumed to be $m_s = 1$), while the latter and the main text use $\kappa_1$ as the overall rate with which the system relaxes back into the ground state $m_s = \pm 1$ sub-manifold. It is then implied by Clua-Provost et al. that this rate is the same for the $m_s = -1$ state, hence the given total lifetime of the IS of $(\kappa_0 + 2 \cdot \kappa_1) = T_{IS}$

The value $k_r$ is not known experimentally. The only estimation that is available is from calculations by Reimers et al. (*28*). Nevertheless, the exact value of this rate is not decisive for the dynamics. Most relevant is that it is significantly smaller than all the other relaxation rates in the system.

The value for $k_e$ is estimated manually by simulation to gain good fitting of the experimental data.

Transient Photoluminescence

Transient PL for various dark periods was conducted in one long pulse train. The sequence is uploaded to the Swabian instruments Pulse Streamer 8/2 device and repeated continuously. PL data is recorded by a combination of an Excelitas single photon detector and the time-correlated single-photon counting module Time Tagger 20 from Swabian instruments. The data is averaged until the signal-to-noise ratio is satisfying.

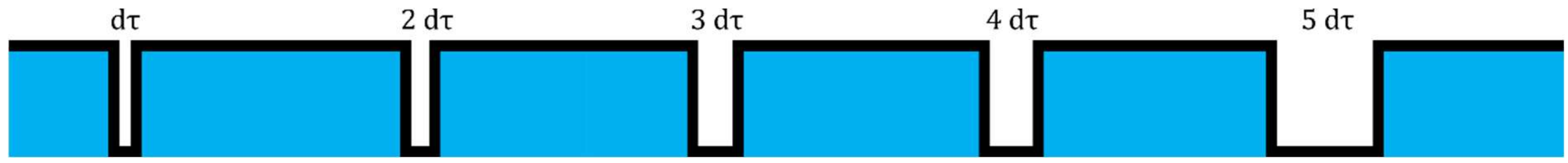

**Fig. S1:** Pulse train schematic for transient PL measurements.

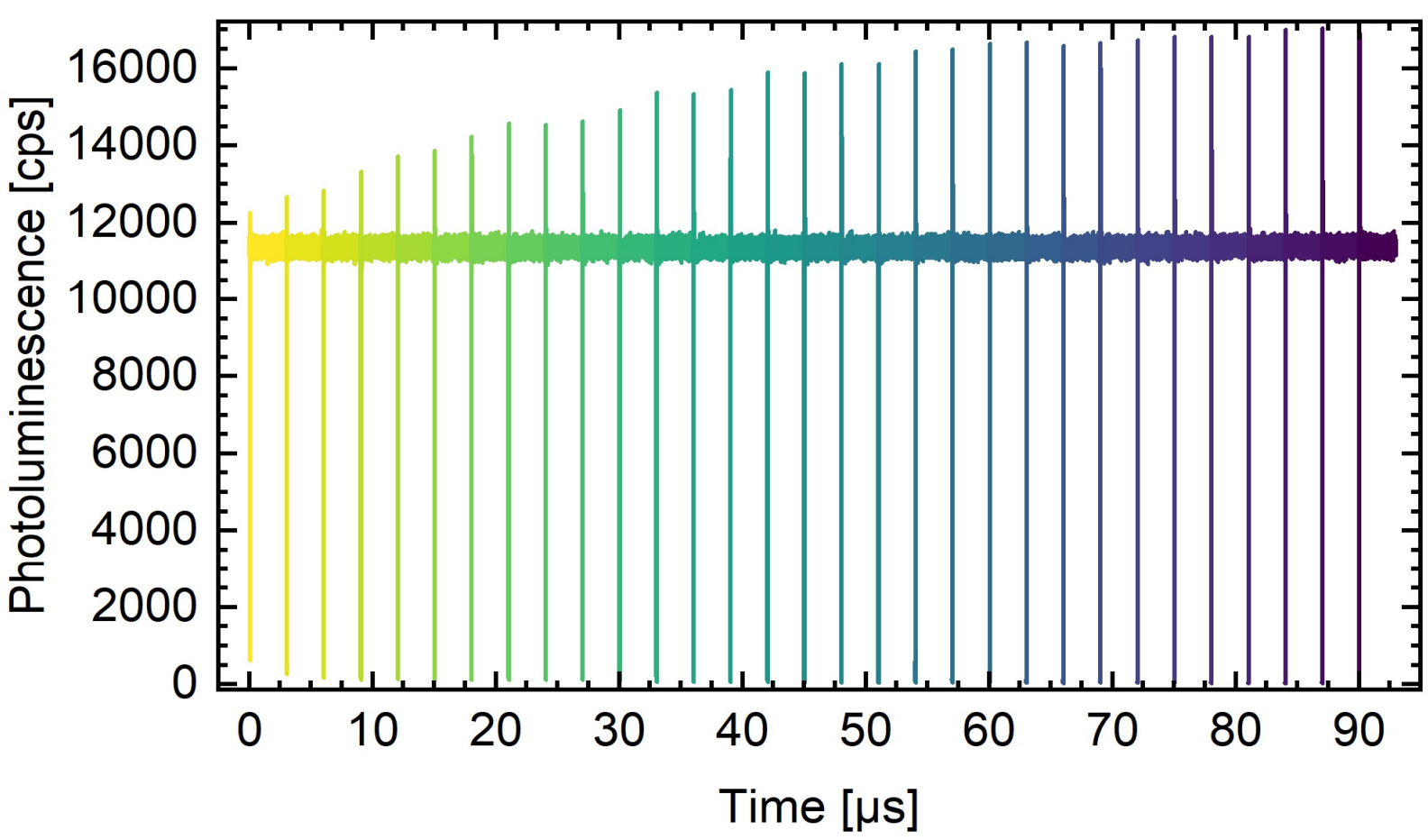

**Fig. S2:** Transient PL trace with increasing dark periods measured as one continuous pulse train. The dark periods (not resolved in this scale) increase from left to right. The increase of the PL overshoot due to repopulation of the GS can clearly be seen. The data reliably shares a common stable state with a shared level in PL intensity.

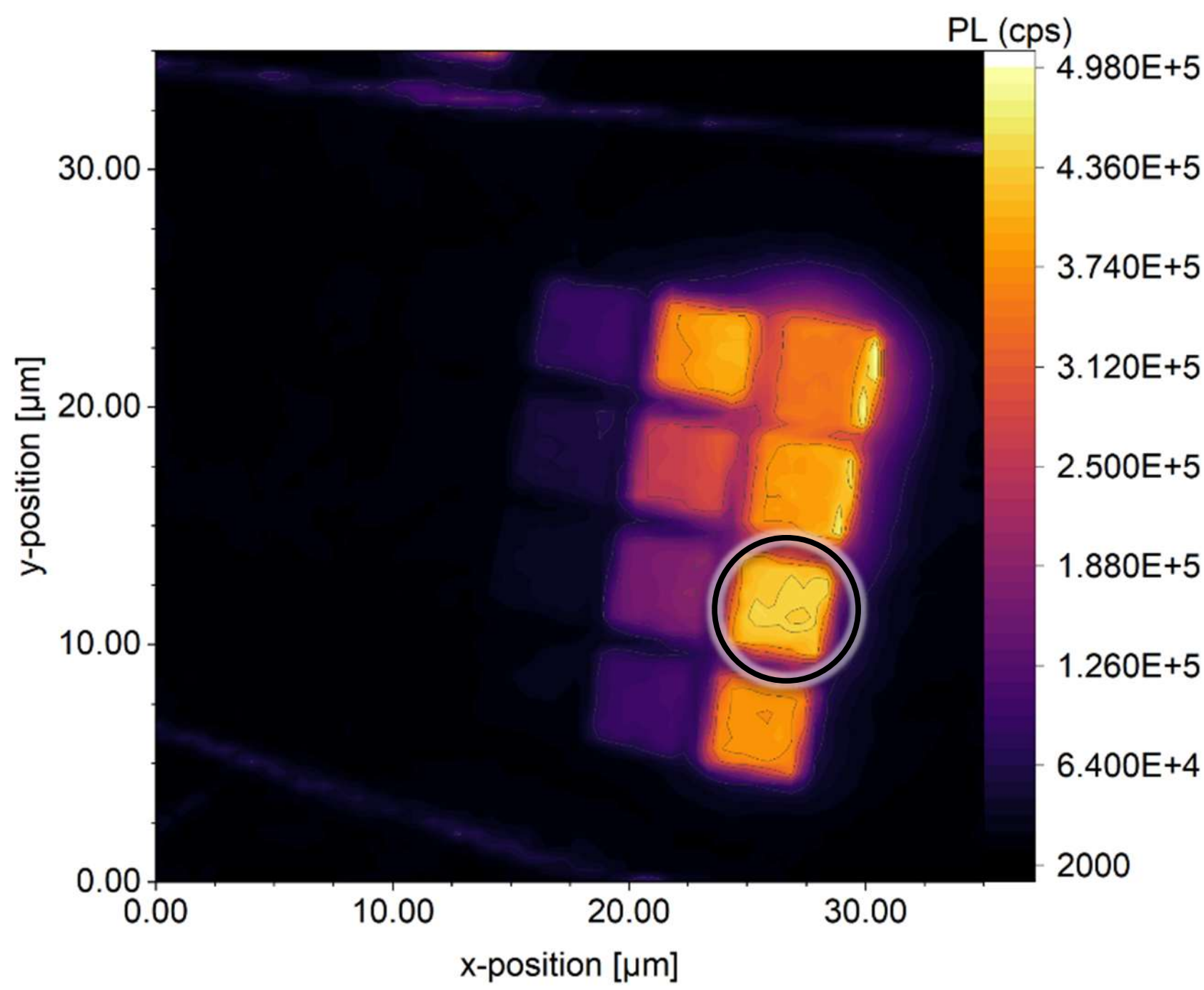

**Fig. S3:** Laterally resolved PL of the hBN sample studied in this work. Measurements were conducted on the irradiated area with the highest PL intensity, highlighted by a circle.

Dependence of $V_B^-$ photoluminescence on laser power

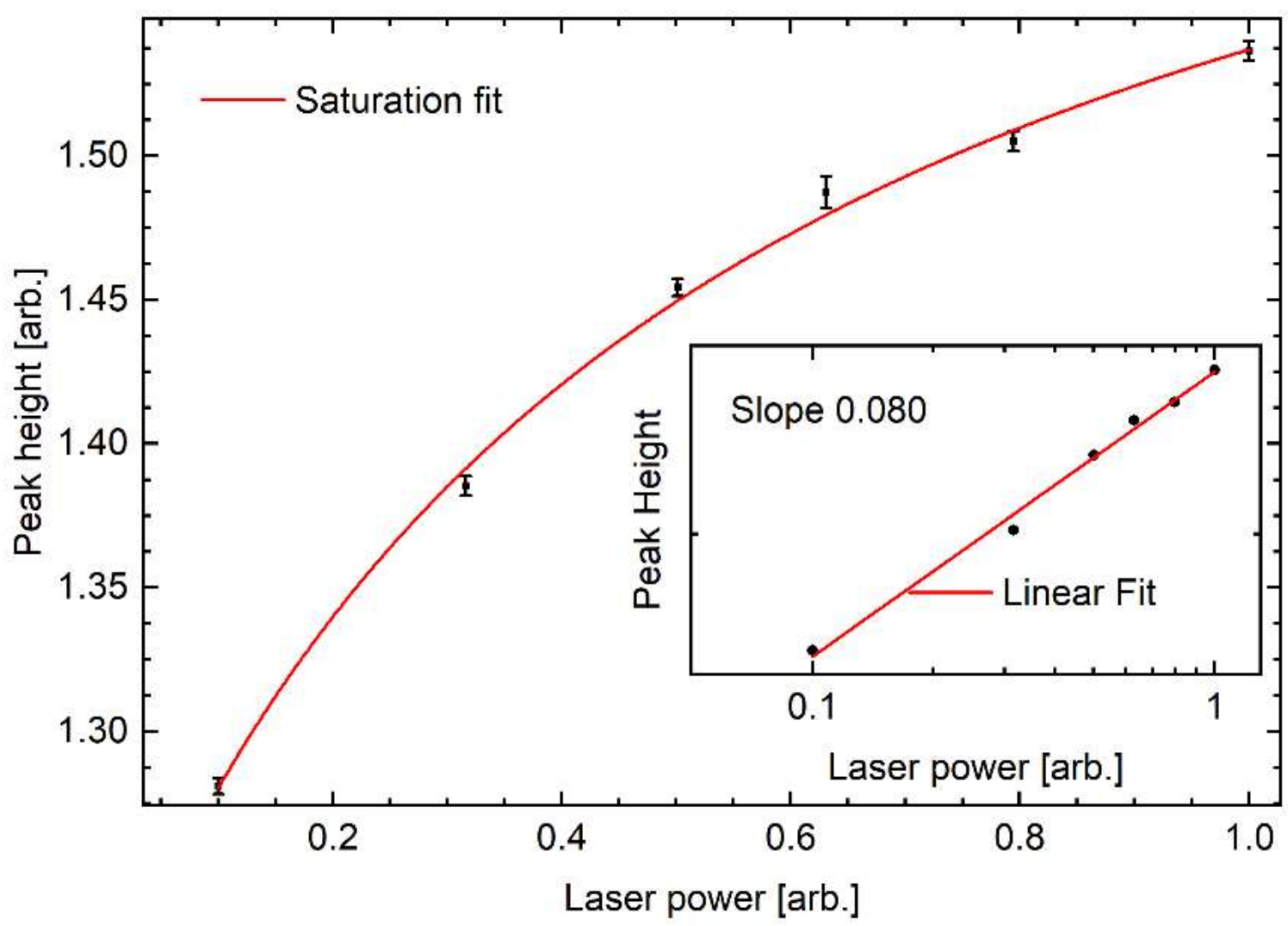

**Fig. S4:** Peak height of the PL overshoot depending on the laser power for a dark period of 150 ns. As can be seen, the peak value begins to saturate, indicating excitation of a large portion of the GS population.

Figure S4 shows the dependence of the maximum overshoot peak height on the laser power for the transients shown in Fig. 2 of the main text. The onset of saturation can be observed, and the following model is fitted to the data:

$$y = y_0 + \frac{h_{\mathrm{sat}}}{1+\frac{P_{½}}{P}}$$

The fit yields $h_{\mathrm{sat}}$=0.52 and $P_{½}$=0.54.

Additional simulations

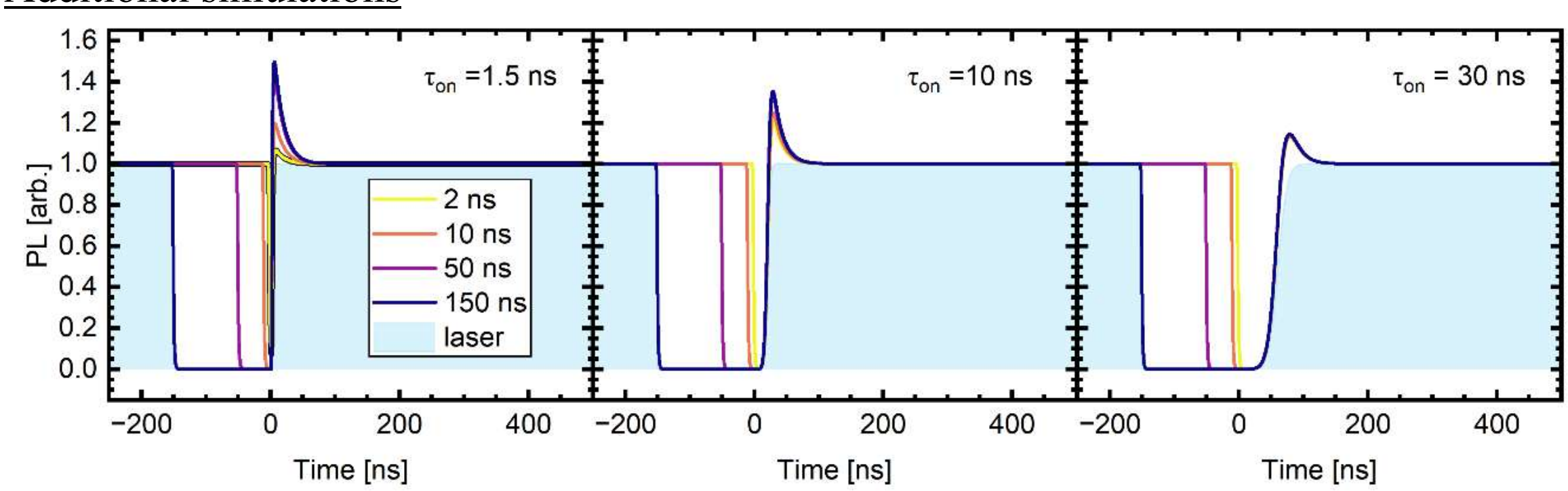

**Fig. S5:** Simulated PL transients for various dark periods. The simulation is repeated for various values of smearing of the rising and falling flanks of the laser excitation. The laser excitation is shown as the blue shaded area for a dark period of 150 ns. The overshoot decreases significantly with softer turn-on of the laser and the dependence on the dark period becomes insignificant.

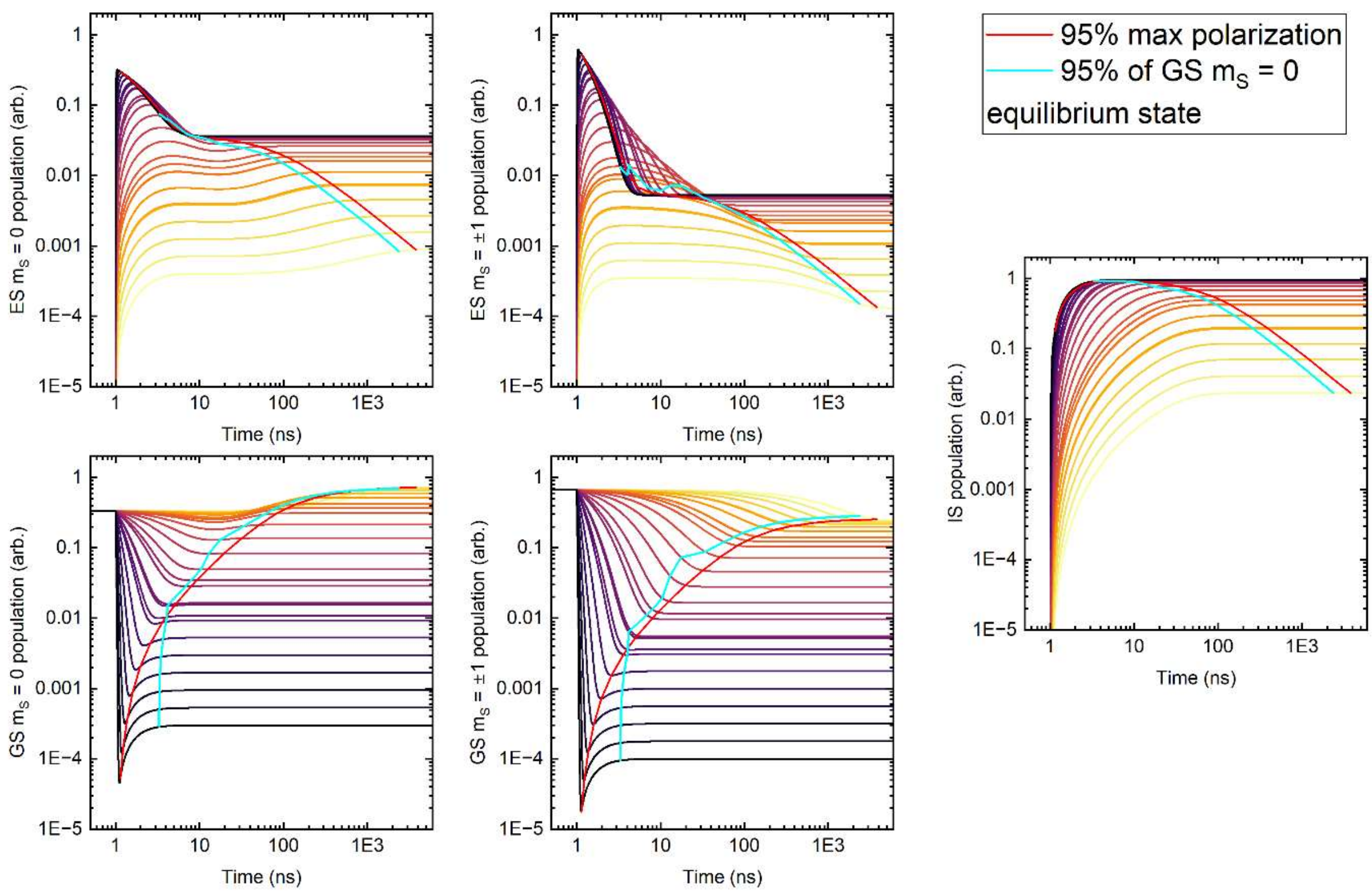

**Fig. S6:** Simulated populations of all sub states for various excitation rates in units of total population. The overlying lines are for two thresholds, namely for equilibrium in the GS $m_S = 0$ state (cyan line) that was also used for Fig. 3c and for the time required to gain 95% of the maximum spin polarization (75%, red line).

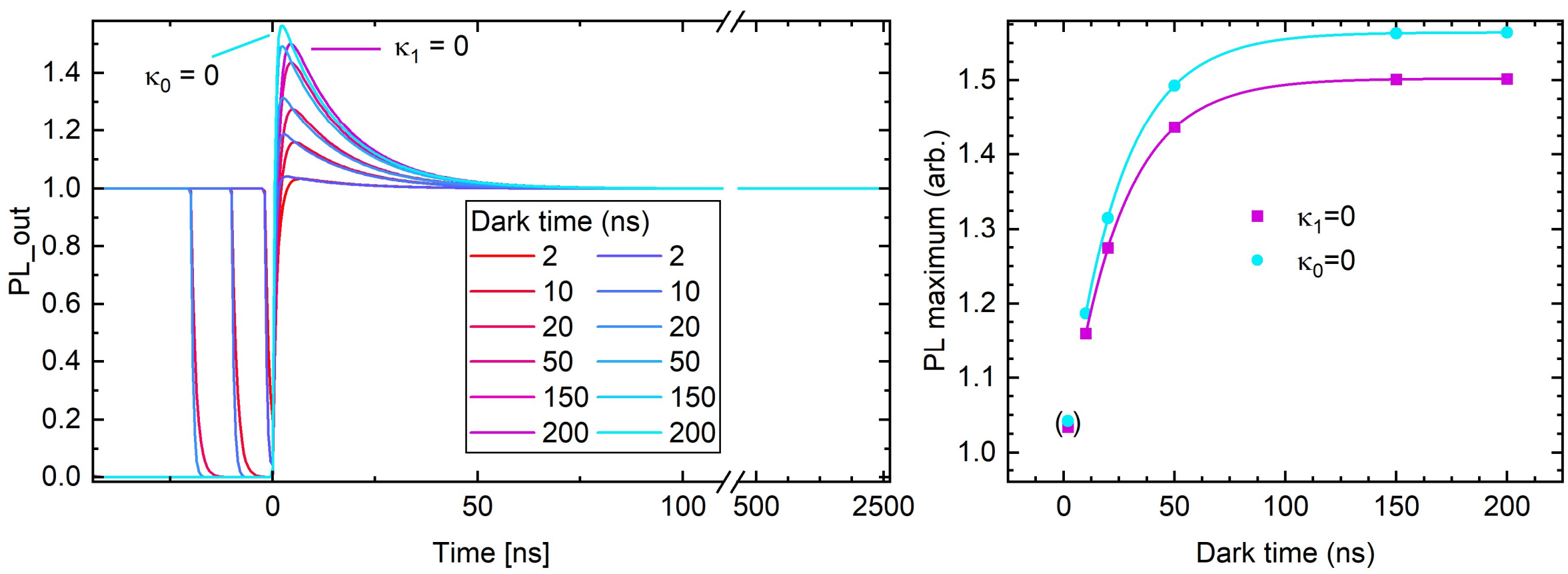

**Fig S7:** Influence of the ratio $\kappa_1/\kappa_0$ on transients. The simulations were conducted for the extreme cases of $\kappa_1 = 0$ and $\kappa_0 = 0$. Left: Transients with the given $\kappa$ set to 0. The $\kappa$ ratio is not experimentally accessible with our method. The transients do not differ significantly and therefore do not change the interpretation of our data. Right: Calculated PL overshoots for the respective $\kappa$ set to 0. The calculated PL maximum follows the same exponential trend with the same IS lifetime. In the fits, the first data point for 2ns dark period is omitted since the relaxation rates are too slow to deplete the ES in time, resulting in an expected deviation from exponential growth.

Laser response measured by the photoluminescence of silicon substrates

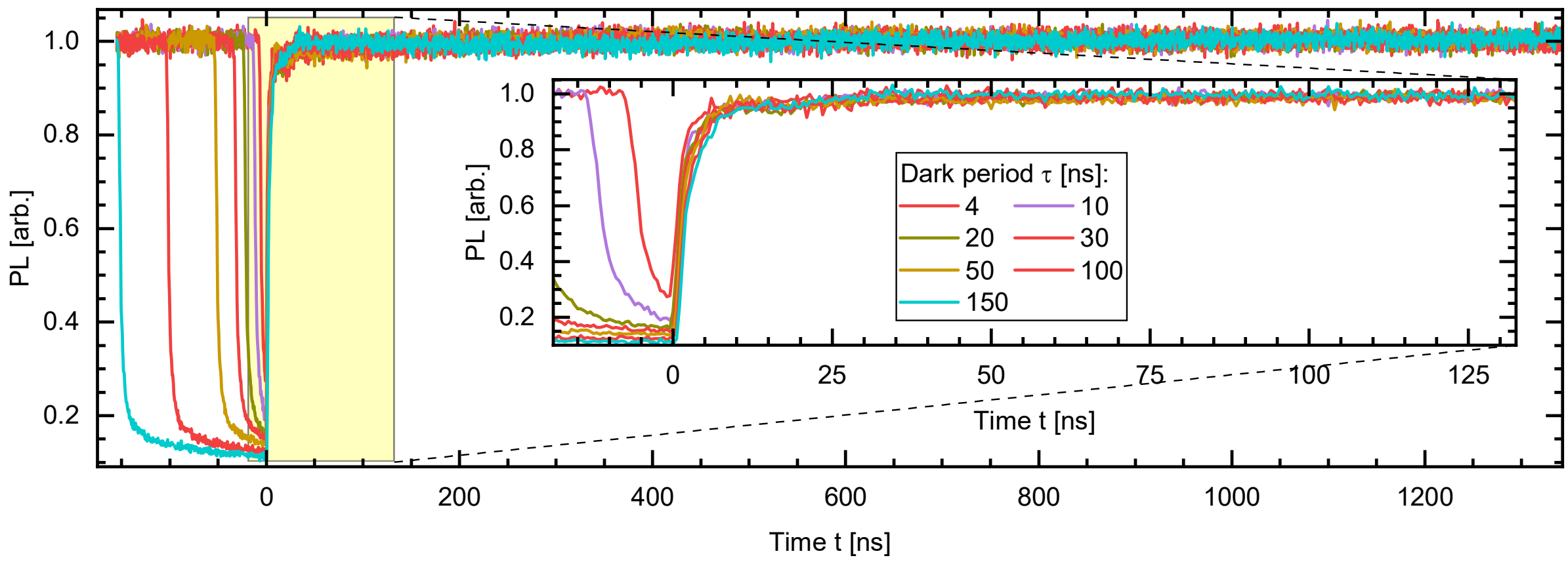

**Fig. S8:** Transient PL of the silicon substrate for several dark periods. No PL overshoot is observed. This shows that the PL overshoot of $V_B^-$ discussed in this paper is not an artifact of the laser on/off modulation and / or instrument response function.

**Code S1**

Python code for simulating the transient photoluminescence with an example for a pulsed laser excitation and dark period of 150ns. Requires the packages numpy and matplotlib

**Data S1**

Data shown in the figure plots of the main text. Each subfigure is listed as a separate sheet. Data in subfigures containing multiple datasets is labeled column-wise. The most left column of each data block contains x-axis values.